# How Much is Enough? An Empirical Test of the Resource Dispersion Hypothesis

Sourabh Biswas[1], Kalyan Ghosh[1], Sumedha Touhid[2], Srijaya Nandi[1], Arpan Bhattacharyya[1], Arunima Bhattacharyya[3], Milisha Das[4], Raktim Paul[5], and Anindita Bhadra[1]*

**Affiliations**

[1]Behaviour and Ecology Lab, Department of Biological Sciences, Indian Institute of Science Education and Research Kolkata, Mohanpur, West Bengal, India

[2]Department of Life Sciences, Presidency University, Kolkata, West Bengal, India

[3]Department of Biotechnology, Maulana Abul Kalam Azad University of Technology, Nadia, West Bengal, India

[4]Department of Microbiology, Kanchrapara College, North 24 Parganas, West Bengal, India

[5]Department of Microbiology, Ramakrishna Mission Vivekananda Centenary College, Rahara, North 24 Parganas, West Bengal, India

sb18rs107@iiserkol.ac.in, kg23rs078@iiserkol.ac.in, sumedhatouhid@gmail.com, sn21rs025@iiserkol.ac.in, ab21rs003@iiserkol.ac.in, arunimabhattacharyya020@gmail.com, milishadas71905@gmail.com, raktimpaul29@gmail.com, abhadra@iiserkol.ac.in

**ORCID ID**

Sourabh Biswas: https://orcid.org/0000-0001-6187-8106

Kalyan Ghosh: https://orcid.org/0000-0002-2602-8912

Sumedha Touhid: https://orcid.org/0009-0006-6572-2081

Srijaya Nandi: https://orcid.org/0000-0002-9462-2374

Arpan Bhattacharyya: https://orcid.org/0000-0003-2621-7868

Arunima Bhattacharyya: https://orcid.org/0009-0008-9723-2372

Milisha Das: https://orcid.org/0009-0000-6688-2446

Raktim Paul: https://orcid.org/0009-0001-6523-2203

Anindita Bhadra: https://orcid.org/0000-0002-3717-9732

***Address for Correspondence**

Behaviour and Ecology Lab, Department of Biological Sciences, Indian Institute of Science Education and Research Kolkata Mohanpur Campus, Mohanpur, Nadia

PIN 741246, West Bengal, India.

Phone no: +91 33 6136 0000 ext 1223

*Corresponding author

E-mail: abhadra@iiserkol.ac.in

**Abstract:**

Free-ranging dogs (*Canis familiaris*) thrive in diverse landscapes, including those heavily modified by humans. This study investigated the influence of resource availability on their spatial ecology across 52 rural and 41 urban sites, comparing urban and rural environments. Census-based surveys were conducted to understand the distribution of dogs and resources, while territory-based observations were carried out across different seasons to capture temporal variability in dog populations and resource availability. Dog and resource density were significantly higher in urban areas, supporting the Resource Dispersion Hypothesis (RDH). Territory size (TS) varied seasonally, decreasing significantly (by 21%) post-mating, likely reflecting shifts in resource demands and distribution. TS was positively correlated with resource heterogeneity, dispersion, patch richness, and male-to-female ratio, but not with group size, which remained stable across seasons and resource gradients. This suggests that while resource availability and sex ratio influence space use, social factors play a key role in shaping group dynamics. These findings highlight the complex interplay between resource availability, social behaviour, and human influences in shaping the spatial ecology of free-ranging dogs and have important implications for their management and disease control, informing targeted interventions such as spay/neuter programs and responsible waste management in both urban and rural landscapes.

**Introduction:**

Group living is a behavioural tendency observed in diverse taxa of animals, from insects to whales, with varied levels of social organization being evident, from loose herd-like structures like in many ungulates to highly organized social systems as in the case of social insects. While group living offers many obvious benefits like security in numbers, better hunting capacity, cooperative brood care, it also has costs like conflict over resources, increased chances of infection and skewed benefits to individuals due to the formation of hierarchies. Cost-benefit trade-offs that ultimately push towards a stable strategy of living in groups eventually lead to the evolution of these social structures (Pyke, 1979).The Resource Dispersion Hypothesis (RDH) (Macdonald, 1983) provides a framework for understanding the optimization process leading to group formation when resources are patchy, i.e., non-homogenously distributed either in space or time. The RDH predicts that territory size should primarily be determined by resource dispersion, while group size would be influenced by resource richness and heterogeneity. According to the RDH, behavioural interactions are not essential for the formation of groups, though the sustainability of such groups would obviously require interactions. Over the last three decades, the RDH has been the focus of multiple studies, gathering support in many cases, but also being challenged in some (Kruuk & Hewson, 1978; Kruuk *et al.*, 1989; Vangen *et al.*, 2001).

Interestingly, studies on carnivores, the group for which RDH was first proposed, have provided mixed support for the framework. While research on red foxes (*Vulpes vulpes*) (Macdonald, 1983; Doncaster & Macdonald, 1997) and spotted hyenas (*Crocuta crocuta*) (Kruuk, 1985) have aligned with the RDH's predictions, some studies on other carnivores, such as European otters (*Lutra lutra*) (Kruuk & Hewson, 1978; Kruuk *et al.*, 1989) and wolverines (*Gulo gulo*) (Vangen *et al.*, 2001), have shown deviations from the expected patterns. These contrasting findings underscore the complexity of factors influencing territoriality and sociality in carnivores, where factors like predation risk, competition and social dynamics may play prominent roles in the determination and maintenance of territories.

Canids display a range of social organization, from living solitarily (Dietz, 1984) or in pairs (Kamler & Macdonald, 2014), to forming well organized packs (Zubiri & Gottelli, 1995; Creel, Mills, & McNutt, 2004). Hence, they have been attractive model systems for testing the RDH. Among canids, free-ranging dogs (FRDs) (*Canis familiaris*) occupy a unique ecological niche, thriving in diverse human-dominated landscapes across the globe (Serpell, 1995; Sen Majumder, Chatterjee, & Bhadra, 2014; Bergström *et al.*, 2020). Pet dogs are completely dependent on human care and support for their survival, but the FRDs thrive as free-living populations, primarily as scavengers dependent on human-generated waste. Though they are capable of hunting, they rarely do so, especially in the urban environment (Majumder *et al.*, 2014; Biswas *et al.*, 2024b). The anthropogenic resources that FRDs rely on for their sustenance are quite fluctuating in quality and quantity, and non-uniform in their distribution in space. FRDs live in social groups and display extensive cooperation-conflict dynamics within and between groups, and present an excellent model system for investigating the interplay between resource availability, territoriality, and social dynamics. Urban areas typically present a higher density and predictability of food resources, potentially leading to smaller territory sizes (Sambo *et al.*, 2018; Thanapongtharm *et al.*, 2021). In contrast, rural areas, characterized by more dispersed and less predictable resources, may necessitate larger territories to ensure access to sufficient food and shelter (Valenzuela & Macdonald, 2002; Warret Rodrigues & Roth, 2023).

Furthermore, the social structure of free-ranging dogs, including group size and composition, is likely influenced by both resource availability and territoriality. In resource-rich environments, larger groups may be sustainable, but in such areas, due to high human density, larger groups might not be favoured by humans. When resources are scarce or highly clumped, the strategy for survival can diverge: smaller groups or solitary living may minimize competition, while larger groups can leverage cooperative hunting (Travis & Slobodchikoff, 1993; Travis, Slobodchikoff, & Keim, 1995). Sen Majumder et al. (2014) provide key insights into the social dynamics of free-ranging dogs, showing that their tendency to forage alone or in groups varies seasonally with specific social needs. During the mating season, adults show a marked increase in opposite-sex associations, while in the non-mating season, juveniles often stay close to adults—suggesting that social groupings serve both reproductive and survival functions. Understanding how these factors interact to shape the social organization of free-ranging dogs is crucial for predicting their ecological roles and potential impacts on local ecosystems.

Despite the growing body of research on free-ranging dogs, a comprehensive understanding of how territory size and group dynamics fluctuate in response to variations in resource distribution and seasonal changes remains elusive. This study investigates how seasonal changes in resource abundance and distribution influence the territory and group sizes of free-ranging dogs (FRDs). We used census surveys across an urban-rural gradient to map dog and resource distribution. Additionally, we conducted territory-based observations across seasons to capture seasonal variability and test the resource dispersion hypothesis, specifically examining how resource patchiness, heterogeneity, and dispersion affect FRD territory and group sizes.

**Methodology:**

**Study sites and subjects**

*Census-based Study*

This study employed a census-based approach to investigate the relationship between resource availability and the spatial ecology of free-ranging dogs across urban and rural environments. Rural and urban areas were classified according to the criteria defined by Ravi (2023), EAC-PM, Govt. of India, ensuring a standardized and consistent classification system. A total of 52 rural sites and 41 urban sites (Figure 1) were surveyed, providing a robust sample size for capturing the variability in dog populations and resource availability across the study region.

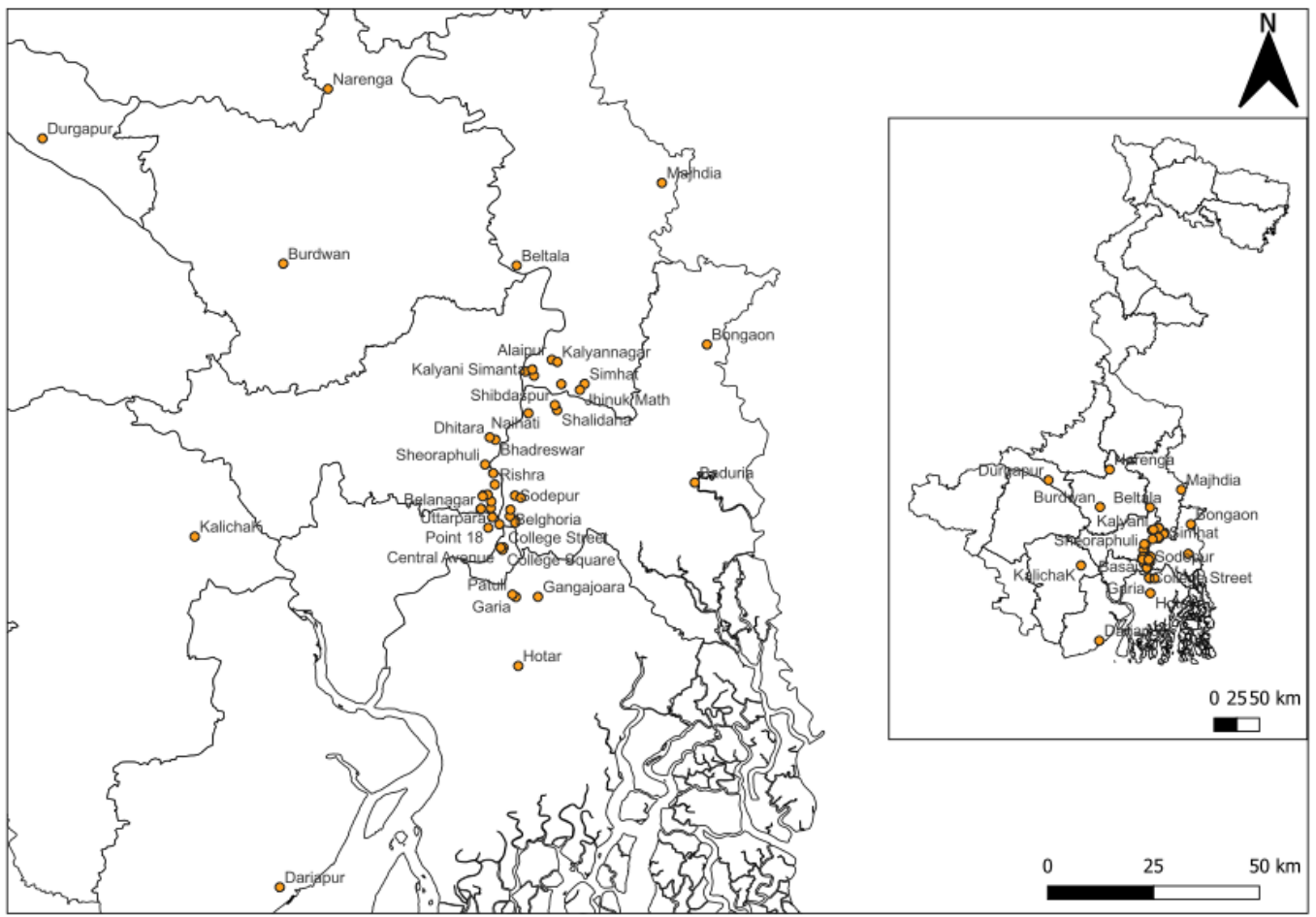

*Figure 1.* Map of West Bengal, India, showing the locations of the study sites. Inset map shows the location of West Bengal within India.

Within each selected location, a polygon (Figure 2 & 3) was randomly drawn on Google My Maps to define the survey area. These polygons ranged in size from 1.29 to 161 hectares and were designed to encompass human settlements and their adjoining accessible areas, ensuring a focus on free-ranging dogs living in close proximity to human habitation. All surveys were conducted between 16:00 and 18:00, a time period when dogs are typically most active and visible (based on prior observations), and daylight conditions facilitate accurate observation and recording. We followed the spot census method established by Sen Majumder et al. (2014).

During each census, several parameters were documented for each free-ranging dog observed, including the time of sighting, sex (determined by observing the genitalia), age class (categorized as pups, juveniles, or adults based on size and genital structures). In order to avoid re-sampling, dogs were recorded on a road only the first time it was traversed during the survey. While every effort was made to conduct thorough surveys within accessible areas, potential biases in dog observations due to variations in observer effort or detectability across different habitats cannot be completely ruled out.

In addition to documenting the presence of dogs, the survey also quantified the type and abundance of resources available to them. The observer walked through the area, marking the positions of various resources, including waste bins, vats and dumps, food stalls, food shops and restaurants, markets, and water sources (such as open taps and open tanks) using GPS following the same protocol as mentioned above (Sen Majumder *et al.*, 2014). Only resources accessible to dogs were considered .Within each polygon, all households were surveyed (with consent) to assess the frequency and type of food provided to free-ranging dogs. This survey aimed to quantify anthropogenic food subsidies, a significant resource for these dogs (Bhattacharjee & Bhadra, 2021). Each resource point within the polygon was assigned a score based on the food provision data, following the methodology described in Bhattacharjee and Bhadra (2021). This information was integrated with the observational data on resource availability to create a comprehensive picture of resource distribution within each survey area.

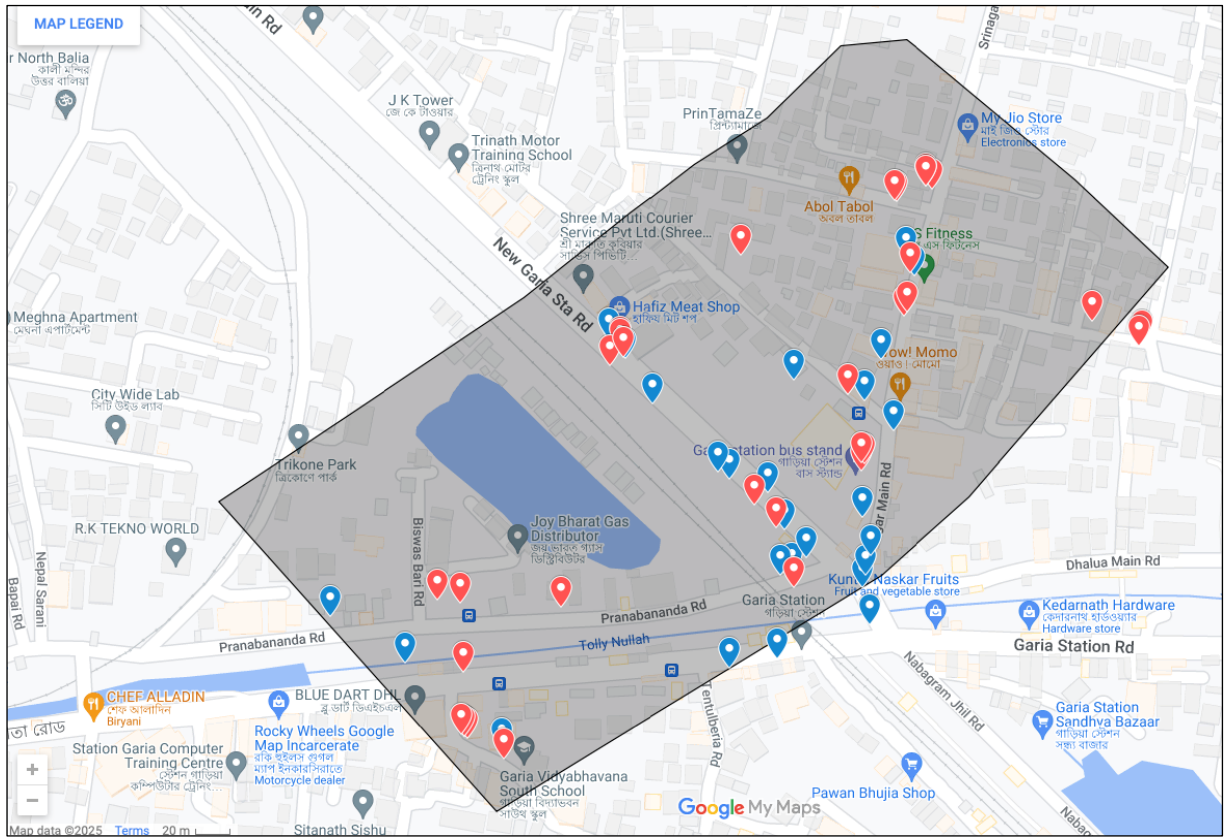

*Figure 2. Spatial distribution of dogs and resources within a sampled urban polygon.*

*This Google My Maps image illustrates the spatial distribution of free-ranging dogs and resource points within a representative urban polygon. Red balloons indicate the locations of individual dogs observed during the census, while blue balloons represent identified resource points within the polygon.*

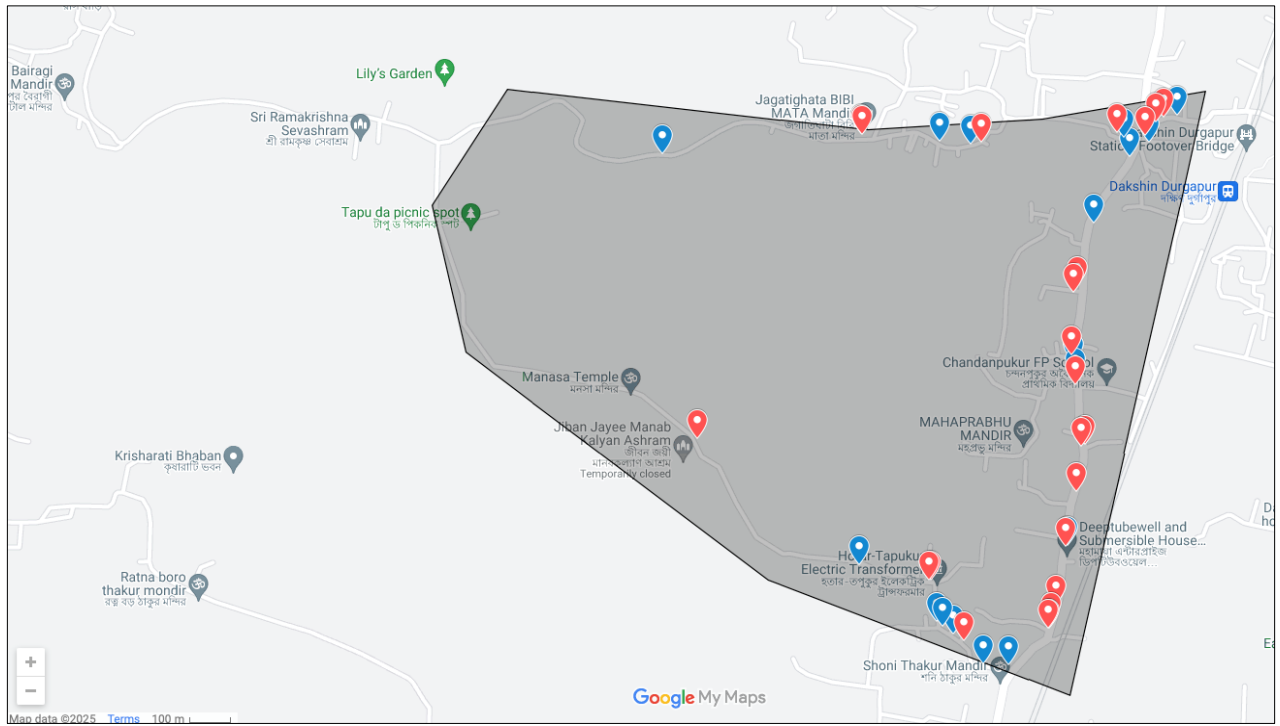

*Figure 3. Spatial distribution of dogs and resources within a sampled rural polygon.*

*This Google My Maps image illustrates the spatial distribution of free-ranging dogs and resource points within a representative rural polygon. Red balloons indicate the locations of individual dogs observed during the census, while blue balloons represent identified resource points within the polygon.*

All surveys were conducted with the utmost respect for the welfare of the dogs and the privacy of residents. Observers maintained a safe distance from the dogs to minimize disturbance, and consent was obtained from households prior to conducting surveys. No dogs were harmed or captured during this study.

***Territory-based study:***

To gain a comprehensive understanding of the spatial and temporal distribution of free-ranging dogs, we conducted a territory-based study across three distinct seasons: the pre-mating season (April to early July), the mating season (mid-July to mid-October), and the post-mating pups emergence season (Late October to March) on 36 dog groups in urban and sub-urban habitat. These seasons were chosen to capture the dynamic interplay between resource fluctuations, territory size, and social behavior in free-ranging dogs, particularly in areas with seasonal fairs or events that temporarily increase resource abundance (Biswas *et al.*, 2024a).

In this study, a territory is defined as an area that an animal or group of animals consistently defends against conspecifics (and occasionally other animals) to secure access to resources and mates. To determine territory boundaries, we conducted behavioral observations of at least 30 hours per group per season, using randomized instantaneous scans and All Occurrences Sessions, following an established protocol (Paul, Majumder, & Bhadra, 2014a). During these observations, we recorded instances of aggressive interactions with adjacent groups, noting the sites of urine marking by individual dogs and the spots actively defended during territorial fights, using GPS. These points were then plotted on Google© My Maps and connected to create polygons representing the territories (Figure 4). While some degree of territory overlap was observed, particularly among related groups, the core areas of exclusive use were clearly unique for each group.

All statistical analyses were conducted in R Studio (version 4.2.0) (R Core Team, 2022). All the GLMs and GLMMs were performed using 'lme4' package (Bates *et al.*, 2015) and 'glmmTMB' package (Bolker, 2019) in R. The final models were selected based on the lowest AIC value. The null and observed models were compared for the models. Model diagnostics were checked using the performance package of R (Lüdecke *et al.*, 2021).

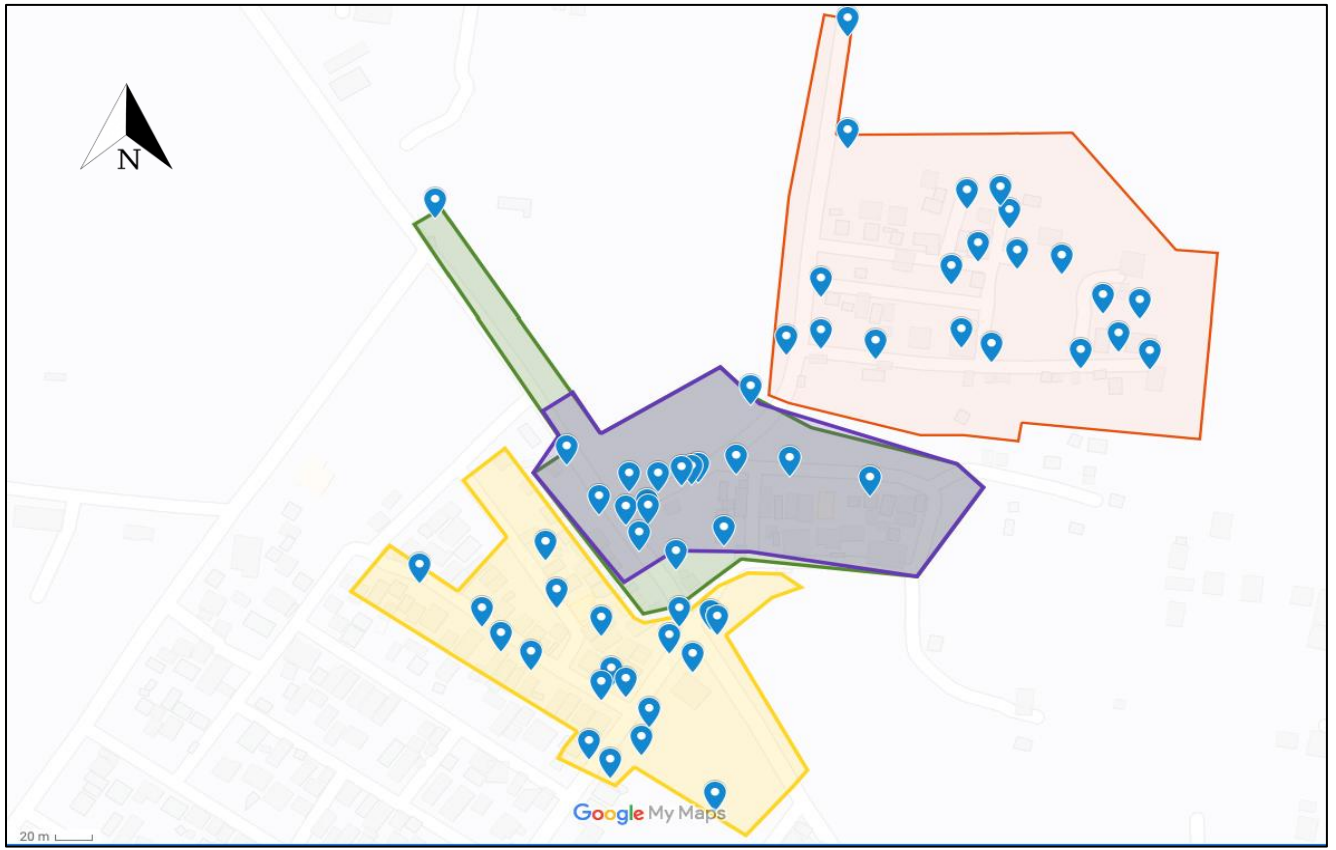

*Figure 4. Territories of three dog groups in Gyaeshpur, Nadia, West Bengal, India. Coloured polygons represent the respective territories of each group. The light green polygon depicts the territory of one group before the mating season. This territory changed after the mating season, and its post-mating boundary is shown by the violet polygon. The territories of the other two groups remained unchanged throughout the study period. Blue balloons indicate resource points within the territories.*

*Census-based Study*

To compare the dog and resource densities between rural and urban areas, a Mann-Whitney U test was employed. A generalized linear model (GLM) with a Gaussian distribution and identity link function was used to assess the relationship between adult dog density (log-transformed for normality) and resource density. Differences in dog density between genders in both urban and rural areas were assessed using a Wilcoxon rank-sum test with continuity correction.

*Territory-based Study*

**Resource mapping**

All resource points within each study territory were identified and categorized and scored (Table 1 following Bhattacharjee and Bhadra (2021). While resources encompass food, water, and shelter, this analysis focuses solely on food resources due to data limitations and the nature of water sources in the study area.

*Table 1: This table presents a scoring system for food sources used by free-ranging dogs, adapted from Bhattacharjee and Bhadra (2021).*

*The "Score" column indicates the quality and quantity of available food, with higher scores representing greater access and nutritional value. Descriptions detail the characteristics of each food source.*

| Sl.No. | Food source | Description | Score |
|---|---|---|---|
| 1 | Meat shop or Fish shop | Open-air shops only, dogs get access to meat or fish scraps. | 8 |
| 2 | Eatery / Restaurant / Hostel | Roadside open-air restaurants/eateries and hostels. Restaurants like 'Subway' and 'McDonald's' were not | 7 |

| | | considered, as free-ranging dogs do not get access to their leftover food. | |
|---|---|---|---|
| 3 | Direct feeding by human or NGO | Direct provision of food by humans (may or may not involve begging) from households, or by an NGO/person who feeds a substantial amount to the dogs. | 6 |
| 4 | Open garbage dump or occasional carcass | $\geq 2$ m2 in size, consisting of wet and dry garbage, including leftover food by humans. It can accommodate several individuals. OR rare or occasional carcass of poultry farm or dead animals. | 5 |
| 5 | Tea shop or Temple | Open-air shops only. Tea shops may also offer bread/biscuits. Temple: Occasionally on festivals. | 4 |
| 6 | Direct feeding by human | Occasional provisioning of food by humans (only by means of begging). | 3 |
| 7 | Grocery / sweet shop/ bakery | All these shops come under one umbrella as they offer direct food provisioning to dogs occasionally. | 2 |
| 8 | Household or eatery dustbins (without cover) | $\leq 0.5$ m2 in size, consisting of wet and dry garbage. It can accommodate up to two individuals at any given time. | 1 |

We quantified resource distribution using three key metrics:

**Patch Richness:** Calculated as the total number of food resource points available within each territory per season.

**Resource Heterogeneity:** Determined by the number of different types of food resources present within each territory, adapted from Bhattacharjee and Bhadra (2021).

**Resource Dispersion:** To calculate this, we first determined the distance of each resource point from all other resource points within the territory using the 'spatstat'(Baddeley & Turner, 2005), 'sf'(Pebesma & Bivand, 2023) and 'sp'(Roger, 2013) packages in R. For each individual resource point, we then calculated an average dispersion score by dividing the sum of distances from that point to all other resource points by the total number of resource points used in the calculation. This resulted in an average dispersion score for each resource point, representing its average distance from all other resources in the territory. These individual resource point dispersion scores were then averaged to obtain the overall resource dispersion score for each territory per season.

Two separate GLMs with Gamma distributions and log link functions were used to assess factors influencing territory size (in hectares) and the total number of adult dogs in a territory. The first model included the total number of adult dogs, resource heterogeneity, patch richness, male-female ratio, and patch dispersion as fixed effects to explain territory size. The second model focused on the total number of adult dogs, with resource heterogeneity, patch richness, male-female ratio, and patch dispersion as predictors.

To evaluate the effect of season on territory size, a generalized linear mixed model (GLMM) with a Gamma distribution and log link function was used, incorporating dog group as a random effect.

Another GLMM with a Gamma distribution and log link function assessed the effects of season on the total number of adult dogs in a territory, also including dog group as a random effect and using 'glmmTMB' package (Bolker, 2019) in R.

Model diagnostics were checked using the 'performance' and 'DHARMa' packages (Hartig & Hartig, 2022) in R. An alpha level of 0.05 was maintained throughout the analysis.

**Results:**

*Census-based Study*

Dog density was significantly higher in urban areas compared to rural areas (Mann-Whitney U test: $Z = -3.0898$, $p < 0.05$), as was resource density ($Z = -5.5765$, $p < 0.0001$). This observation supports the premise that resource availability influences dog distribution. Further reinforcing this relationship, a GLM confirmed a significant positive correlation between resource density and adult dog density ($t = 4.043$, $p < 0.0001$). We found no significant difference between the density of male and female dogs in either urban (Wilcoxon rank-sum test: $W = 701$, $p > 0.05$) or rural ($W = 398$, $p > 0.05$) areas.

*Table 2: Correlation between dog density (per hectare) and resource density (per hectare)*
*GLM formula = dog density per hectare ~ resource density per hectare*

| Term | Estimate | Std. Error | t value | Pr(>t) |
| --- | --- | --- | --- | --- |
| (Intercept) | 1.19 | 0.17 | 6.96 | < 0.001 |
| Resource density | 0.06 | 0.02 | 4.04 | < 0.001 |

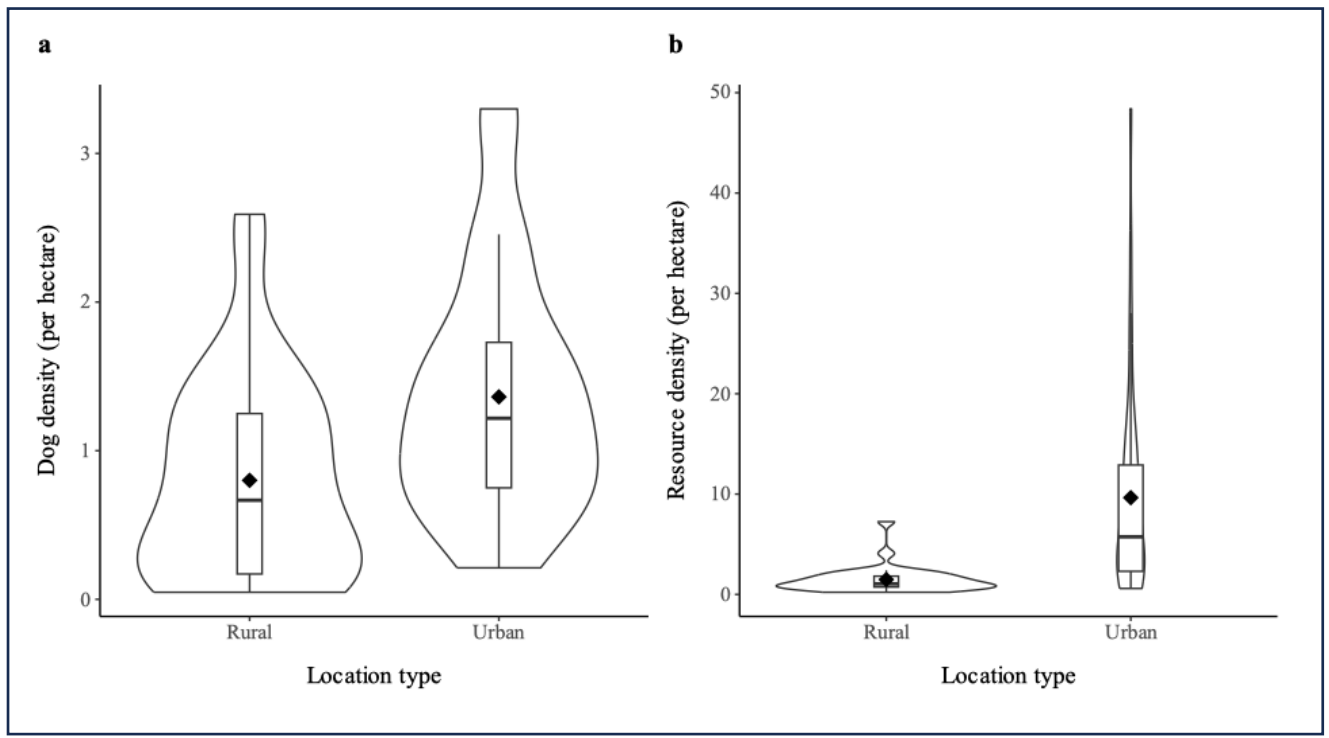

*Figure 5. Dog and resource density across rural and urban locations.*

*(a) Dog density (individuals per hectare) and (b) resource density (resource score per hectare) in rural and urban locations. Violin plots with embedded boxplots illustrate the distribution of densities in each location. Black diamonds within the boxplots represent the mean density for each location.*

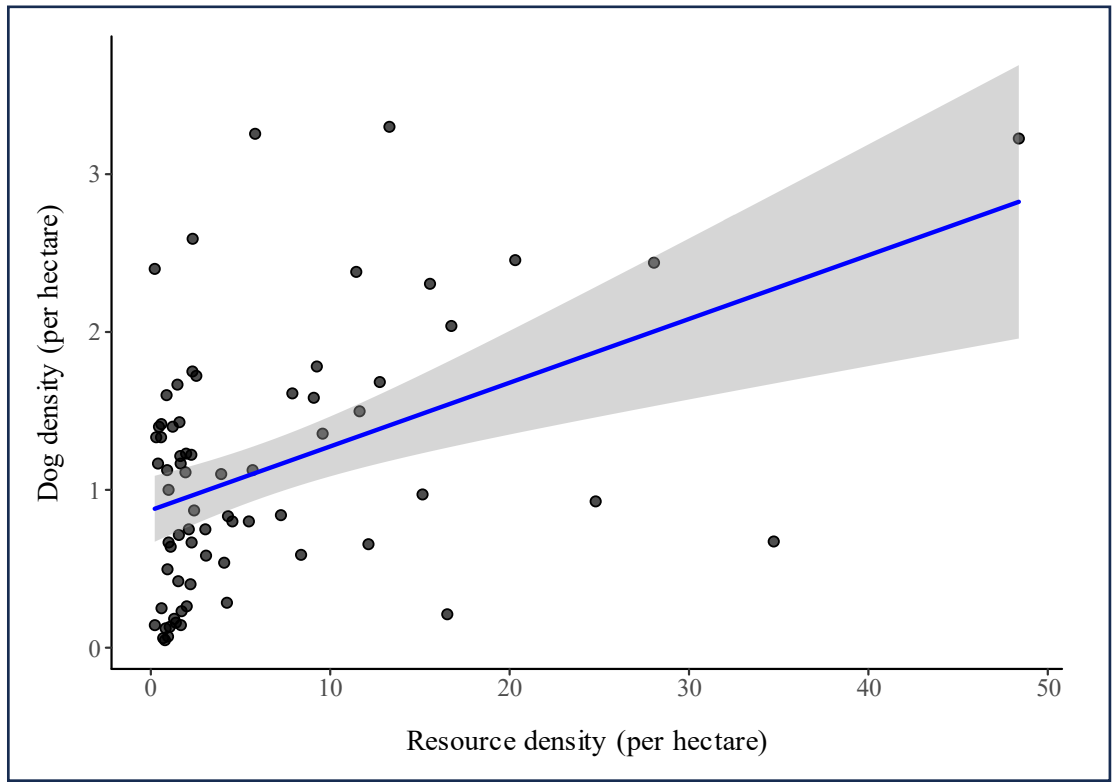

*Figure 6. Relationship between resource density and adult dog density.*

*This scatter plot shows the positive correlation between resource density (units/hectare) and adult dog density (individuals/hectare), with each point representing a location. The blue line depicts the linear regression, and the grey area represents its confidence interval, illustrating the uncertainty in the estimated relationship.*

*Territory-based Study*

Territory size, which ranged from 0.101 to 7.909 hectares (mean = 1.86 ± 1.66 hectares), was found to be dynamic across seasons. A GLMM analysis revealed significant variation in territory size across the three seasons. While territory sizes during the pre-mating (1.89 ± 1.67 hectares) and mating (1.93 ± 1.6 hectares) seasons were comparable ($p > 0.05$), a significant reduction was observed in the post-mating season (1.79 ± 1.57 hectares; z = -2.599, p < 0.01) compared to mating season.

*Table 3: Impact of season on the territory size of free-ranging dog group*

*GLM formula = Territory size ~ Season (Pre-mating/Mating/Post-mating)*

| Term | Estimate | Std. Error | z value | Pr(>z) |
|---|---|---|---|---|
| (Intercept) | 0.33 | 0.18 | 1.84 | > 0.05 |
| Season: Post mating | -0.16 | 0.06 | -2.59 | < 0.05 |
| Season: Pre mating | -0.04 | 0.06 | -0.70 | > 0.05 |

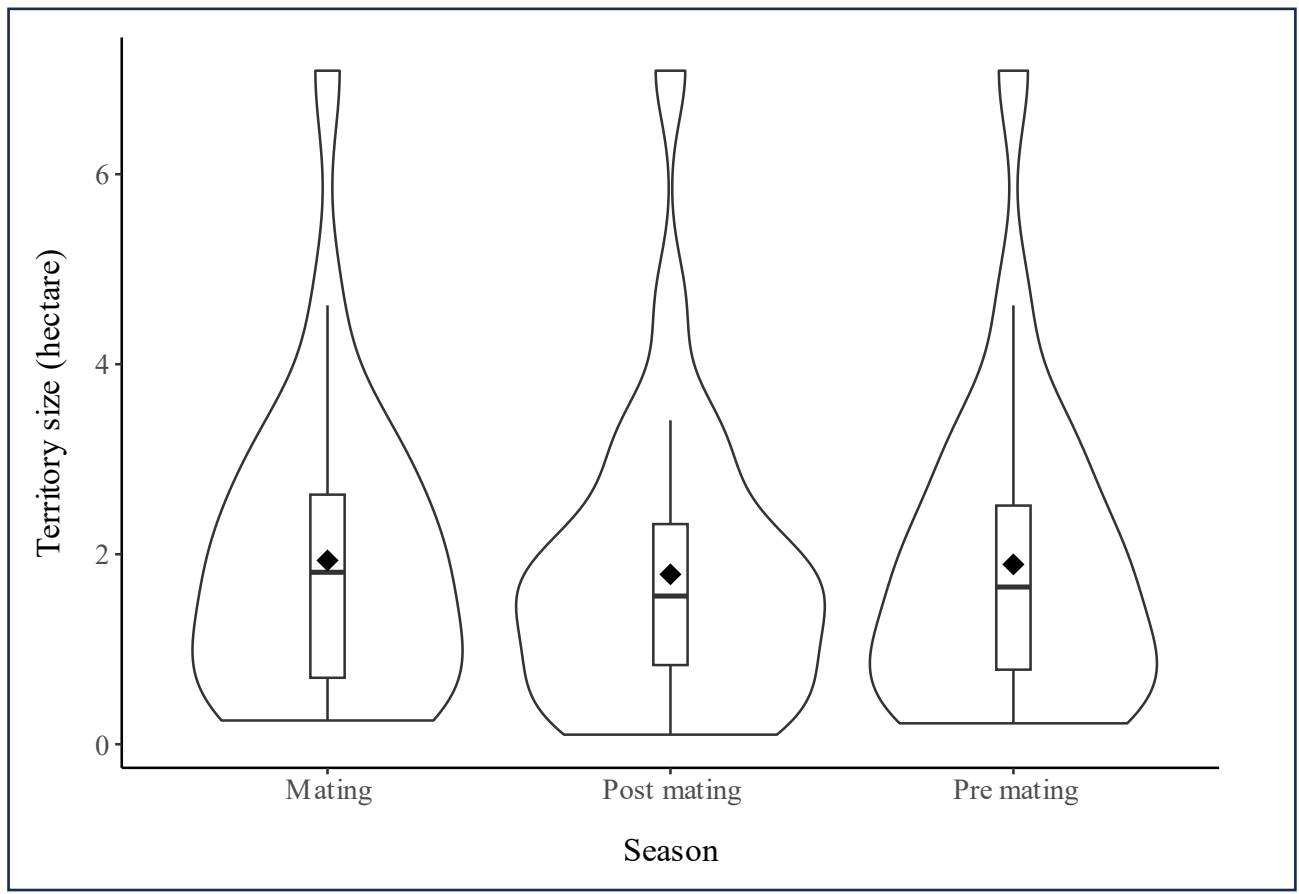

*Figure 3. Seasonal variation in dog group territory size.*

*This figure compares the territory size (hectares) of dog groups across three distinct reproductive seasons: pre-mating, mating, and post-mating. Violin plots with embedded boxplots illustrate the distribution of territory sizes in each season. Black diamonds within the boxplots represent the mean territory size for each season.*

Further investigation into the factors influencing territory size revealed that resource heterogeneity (t = 2.076, p < 0.05), patch richness (t = 3.271, p < 0.01), and resource dispersion (t = 6.740, p < 0.0001) were all positively associated with territory size (GLM analysis). Additionally, the male-female ratio within a group also showed a positive association with territory size (t = 2.085, p < 0.05). However, the total number of adult dogs did not significantly affect territory size (p > 0.05).

*Table 4: Results of a Generalized Linear Model (GLM) predicting territory size in free-ranging dogs.*

*The model examines the influence of group size, resource heterogeneity, patch richness, resource dispersion, and male-female ratio on the size of dog territories.*

*GLM formula: Territory Size ~ Group size + Resource Heterogeneity + Patch Richness + Resource Dispersion + Male-Female Ratio*

| Term | Estimate | Std. Error | t value | Pr(>t) |
| --- | --- | --- | --- | --- |
| (Intercept) | -2.78 | 0.51 | -5.42 | < 0.001 |
| No of dogs | -0.00 | 0.05 | -0.01 | > 0.05 |
| heterogeneity | 0.10 | 0.05 | 2.08 | 0.05 |

| Patch richness | 0.0411817 | 0.0125918 | 3.271 | > 0.05 |
| Male-female ratio | 0.3914299 | 0.1877555 | 2.085 | 0.05 |
| Resource dispersion | 0.0205599 | 0.0030506 | 6.740 | < 0.001 |

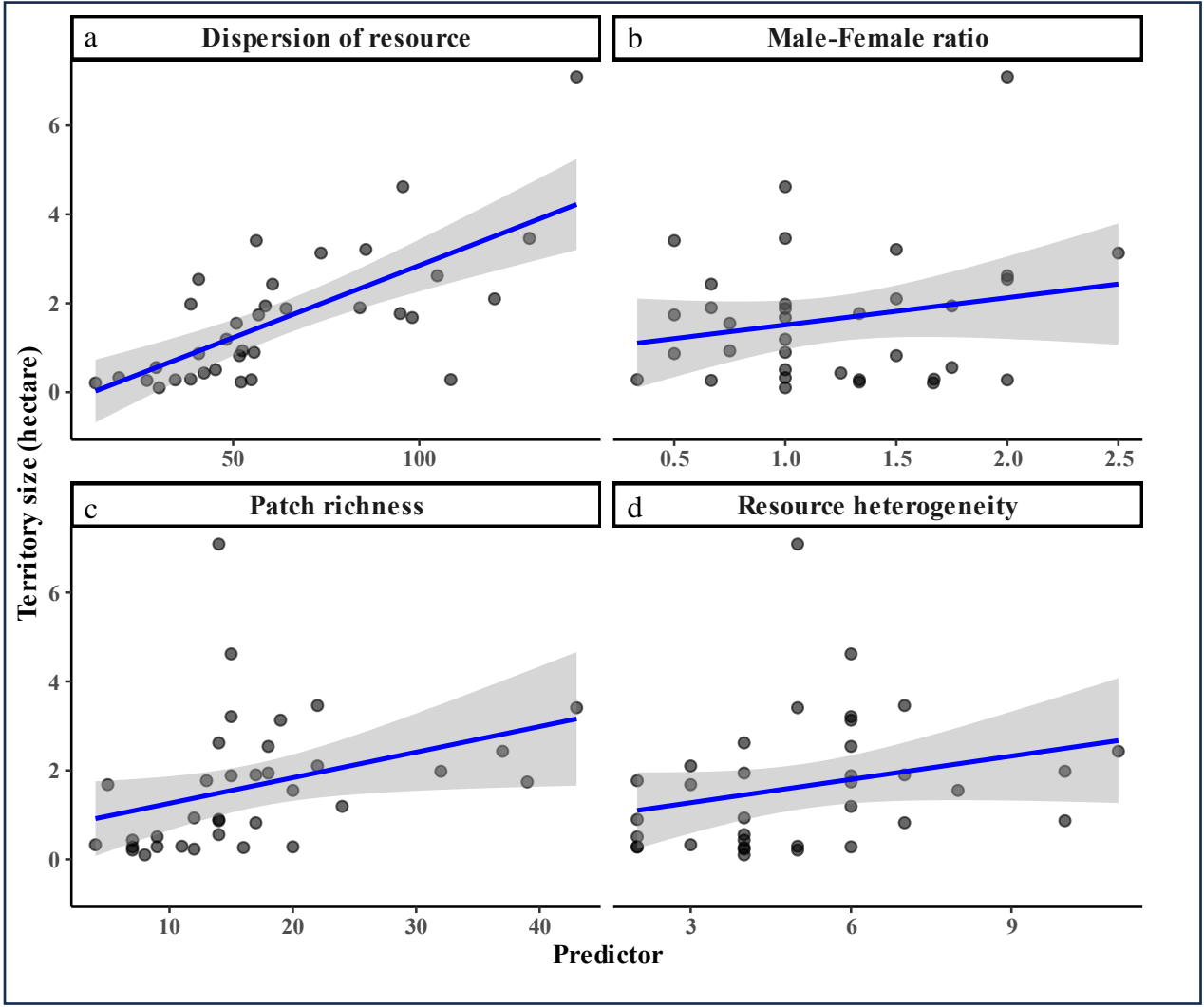

*Figure 4. Factors influencing dog group territory size.*

*This figure presents four scatter plots examining the relationship between territory size (hectares) and various ecological factors: (a) resource dispersion, (b) male-female ratio within the group, (c) patch richness, and (d) resource heterogeneity. In each plot, the blue line represents the linear regression, and the grey area depicts its confidence interval, illustrating the uncertainty associated with the estimated relationship.*

Finally, we examined whether resource availability and seasonality influenced the number of adult dogs within a territory. Neither season (GLMM analysis; $p > 0.05$) nor resource heterogeneity, resource dispersion, or patch richness (GLM analysis; $p > 0.05$) significantly affected the number of adult dogs in a territory.

*Table 5: Results of a Generalized Linear Model (GLM) predicting group size in free-ranging dogs.*

*The model examines the influence of resource heterogeneity, patch richness, resource dispersion, and male-female ratio on the size of dog group.*

*GLM formula: Group size ~ Resource Heterogeneity + Patch Richness + Resource Dispersion + Male-Female Ratio*

| Term | Estimate | Std. Error | t value | Pr(>t) |
|---|---|---|---|---|
| (Intercept) | 1.86 | 0.22 | 8.37 | < 0.001 |
| heterogeneity | -0.00 | 0.02 | -0.26 | > 0.05 |
| Patch richness | 0.00 | 0.00 | 0.25 | > 0.05 |
| Male-female ratio | 0.12 | 0.10 | 1.21 | > 0.05 |
| Resource dispersion | -0.00 | 0.00 | -1.57 | > 0.05 |

**Discussion:**

Free-ranging dogs thrive in a wide range of habitats, including those heavily modified by human activities. This presents a unique opportunity to investigate the interplay between resource availability and the spatial ecology of a socially complex carnivore. This study aimed to unravel this intricate relationship by combining census data with detailed territory analyses, seeking to understand how resource dispersion shapes both population density and individual space use.

Our findings reveal interesting aspects of the complex relationship between free-ranging dogs and their human-dominated environment. As obligate scavengers (Sarkar, Sau, & Bhadra, 2019) intrinsically linked to human societies, they offer a powerful lens through which to examine the complexities of human-animal interactions. These dogs have mastered the art of navigating anthropogenic food sources, their lives intricately woven into the fabric of human activity (Biswas *et al.*, 2024a). Their resource landscape is not defined by naturally occurring prey or vegetation, but by the unpredictable distribution of human-generated waste and deliberate provisioning by humans (Sen Majumder et al. 2014). This reliance on human-derived resources has profound implications for their spatial ecology and social behaviour.

Our census-based approach revealed higher dog and resource densities in urban environments compared to their rural counterparts, with a clear positive correlation between the two. This observation strongly supports the Resource Dispersion Hypothesis (RDH) (Macdonald, 1983), which posits that animal distributions are intrinsically linked to the availability of essential resources. The abundance of anthropogenic resources characteristic of urban areas, such as discarded food waste and deliberate feeding by residents, likely underpins the observed higher dog densities. This aligns with findings from other studies documenting higher dog densities in resource-rich urban areas (Sambo *et al.*, 2018; Thanapongtharm *et al.*, 2021). While overall dog abundance was higher in urban areas, we found no significant differences in the distribution of the two sexes across habitats. This suggests that resource availability influences overall population density, rather than driving sex-specific biases in population structure, as noted by Warembourg et al. (2021).

Moving beyond population-level patterns, we investigated the influence of resource dispersion on individual space use by examining territory size (TS) across seasons and in relation to resource attributes. Our findings demonstrate that TS is quite dynamic, it decreases during the post-mating season as compared to the mating season, while remaining relatively stable in the pre-mating period. This seasonal variation in TS likely reflects shifts in resource demands and distribution, potentially linked to breeding activities and the energetic costs of reproduction (Jenner, Groombridge, & Funk, 2011; Marneweck *et al.*, 2019), as well as mating interests.

However, this dynamic is further complicated by the unique fluctuations of human-generated food waste (Bhattacharjee & Bhadra, 2021), a key resource for free-ranging dogs. Periods of abundance, perhaps following festivals or holidays, might allow dogs to contract their territories, capitalizing on readily available food (Biswas *et al.*, 2024a). Conversely, leaner times could necessitate expanding their ranges to encompass a wider array of potential food sources. This flexibility in space use underscores their adaptability and resilience in a fluctuating resource landscape.

Our results further support the Resource Dispersion Hypothesis (RDH), indicating that territory size (TS) is positively correlated with resource heterogeneity, dispersion, and patch richness, but not with group size. This suggests that while dogs may expand their territories to encompass a wider array of resources, the sizes of their social groups remain stable regardless of resource availability. This suggests an adaptive understanding of the spatial distribution of human-derived resources. Free-ranging dogs are not simply responding to overall abundance, but also to the spatial configuration of these resources, demonstrating a capacity for sophisticated spatial memory and decision-making within the human-modified environment. As resource patches become more dispersed, individuals may need to increase their TS to secure sufficient resources, but this expansion does not necessarily translate to larger group sizes (Valeix, Loveridge, & Macdonald, 2012b; Erlandsson *et al.*, 2022). This decoupling of TS and group size is a crucial observation and aligns with the predictions of the RDH, as well as empirical findings in other social carnivores, such as wolves (Kittle *et al.*, 2015). In fact, we speculate that in areas where a large number of dogs are provided ample food in bulk by dedicated human feeders, the territory sizes might be much reduced, leading to a high density of dogs in such pockets. Future studies need to be conducted to test this prediction.

The positive association between TS and male-to-female ratio suggests that increased male-male competition for mates is another key factor influencing territory size. This could be due to males needing to cover more ground to find mates (Clutton-Brock, 1995), defend resource-rich areas that attract females (Alatalo, Lundberg, & Glynn, 1986), or engage in more aggressive territorial defence (David, 1970; Mitani & Rodman, 1979). For instance, studies in red deer have demonstrated that males with larger territories have greater mating success (Clutton-Brock, Guinness, & Albon, 1982).

Taken together, these findings underscore the complex interplay between resource availability and social dynamics in shaping the spatial organization of free-ranging dog populations. The stability of group size, despite variations in resource availability, sex ratio, and territory size, suggests that social factors, such as dominance hierarchies, competition for mates, and the benefits of cooperative breeding (Paul, Majumder, & Bhadra, 2014b; Paul & Bhadra, 2018), play a crucial role in constraining group size. This is consistent with the work of Nel et al. (2013) and Valeix et al. (2012b), who highlighted the role of social factors, such as intra-group competition and the costs and benefits of group living, in constraining group size. Further research is needed to tease apart the relative importance of these social factors in different resource contexts and to fully elucidate their interplay in shaping the spatial ecology of free-ranging dogs.

This intimate association with humans exposes free-ranging dogs to the full spectrum of human behaviour (Bhattacharjee, Sau, & Bhadra, 2018; Bhattacharjee *et al.*, 2021). While positive interactions, such as deliberate feeding and active begging from humans represent a significant source of sustenance for these dogs (Bhattacharjee *et al.*, 2018, 2021), this reliance on human provisioning also makes them vulnerable to negative human actions. Sadly, humans are a major cause of high early life mortality in free-ranging dogs (Paul *et al.*, 2016).

The implications of these findings for dog population management and disease control are significant. In urban areas, where dog densities are high and resources are concentrated, targeted interventions may be needed to manage populations and reduce the risk of disease transmission. This could involve strategies such as spay/neuter programs, public education campaigns promoting responsible feeding practices, and improved waste management initiatives to limit the availability of anthropogenic food sources. In contrast, management efforts in rural areas may need to focus on different challenges, such as reducing conflict with livestock and wildlife through community-based dog vaccination programs.

While the RDH provides a strong framework for interpreting our results, it is important to acknowledge that other factors, such as social structure, inter-group competition, and human intervention, could also contribute to the observed patterns. Future research should aim to disentangle the relative importance of these factors in shaping the spatial ecology of free-ranging dogs.

**Acknowledgements:**

The authors would like to acknowledge the Indian Institute of Science Education and Research Kolkata for providing infrastructural support.

**Funding:**

SB would like to thank the University Grants Commission, India for providing him doctoral fellowship. The project was partially funded by the Janaki Ammal Award grant BT/HRD/NBA-NWB/39/2020-21 (YC-1), to AB by the Department of Biotechnology, India.

**Author contributions**

SB, KG, ST, SN, ArB, AruB, MD and RP carried out the field work. SB curated the data, carried out the data analysis and wrote the manuscript. SB and AnB conceived the idea. AnB supervised the work, acquired funding, reviewed and edited the manuscript.

**Data availability statement**

Data supporting the results will be archived.

**Conflict of Interest Information**

Authors have no conflict of interest.